\documentclass[12pt]{iopart}

\usepackage{harvard} 
\usepackage{nicefrac} 
\usepackage{graphicx}
\usepackage{epstopdf}
\usepackage{booktabs}
\begin{document}


\title[Low Dimensional Embedding of Climate Data for Radio Astronomical Site Testing]{Low Dimensional Embedding of Climate Data for Radio Astronomical Site Testing in the Colombian Andes}

\author{Germ¬\'an Chaparro Molano$^{1}$}
\address{$^1$Grupo de Simulaci\'on, An\'alisis y Modelado, Vicerrector\'ia de Investigaci\'on, Universidad ECCI, Bogot\'a, Colombia}
\ead{gchaparrom@ecci.edu.co}
\author{Oscar Leonardo Ram\'irez Su\'arez$^{1}$}
\address{$^1$Grupo de Simulaci\'on, An\'alisis y Modelado, Vicerrector\'ia de Investigaci\'on, Universidad ECCI, Bogot\'a, Colombia}
\ead{oramirezs@ecci.edu.co}
\author{Oscar Restrepo$^{1,2}$}
\address{$^1$Grupo de Simulaci\'on, An\'alisis y Modelado, Vicerrector\'ia de Investigaci\'on, Universidad ECCI, Bogot\'a, Colombia}
\address{$^2$Radio Astronomy Instrumentation Group, Universidad de Chile, Santiago de Chile, Chile}
\ead{orestrepog@ecci.edu.co}
\author{Alexander Mart\'inez$^{1,3}$}
\address{$^1$Grupo de Simulaci\'on, An\'alisis y Modelado, Vicerrector\'ia de Investigaci\'on, Universidad ECCI, Bogot\'a, Colombia}
\address{$^3$Instituto de Hidrolog\'ia, Meteorolog\'ia y Estudios Ambientales, Bogot\'a, Colombia}
\vspace{10pt}
\begin{indented}
\item[]
\end{indented}

\begin{abstract}
We set out to evaluate the potential of the Colombian Andes for millimeter-wave astronomical observations. Previous studies for astronomical site testing in this region have suggested that nighttime humidity and cloud cover conditions make most sites unsuitable for professional visible-light observations. Millimeter observations can be done during the day, but require that the precipitable water vapor column above a site stays below $\sim$10 mm. Due to a lack of direct radiometric or radiosonde measurements, we present a method for correlating climate data from weather stations to sites with a low precipitable water vapor column. We use unsupervised learning techniques to low-dimensionally embed climate data (precipitation, rain days, relative humidity, and sunshine duration) in order to group together stations with similar long-term climate behavior. The data were taken over a period of 30 years by 2046 weather stations across the Colombian territory. We find 6 regions with unusually dry, clear-sky conditions, ranging in elevations from 2200 to 3800 masl. We evaluate the suitability of each region using a quality index derived from a Bayesian probabilistic analysis of the station type and elevation distributions. Two of these regions show a high probability of having an exceptionally low precipitable water vapor column. We compared our results with global precipitable water vapor maps and find a plausible geographical correlation with regions with low water vapor columns ($\sim10$ mm) at an accuracy of $\sim20$ km. Our methods can be applied to similar datasets taken in other countries as a first step toward astronomical site evaluation.
\end{abstract}

%
%
\submitto{PASP}
%
%
%


\section{Introduction}

The development of astronomical instrumentation technology in the 0.1-2 THz range has been rapidly growing in recent years. Another recent surge in interest in technology at this frequency range has appeared recently due to the saturation of the ``classical radio window'' in telecommunications  \cite{newera}. For this reason, atmospheric models for absorption of THz photons \cite{rosenkranz,lababs} and artifact removal models \cite{removal} have been developed. Countries near the Equator face a challenge in using this frequency band for astronomical observations and telecommunications using terabit satellite links  \cite{suen2016} due to the presence of a tropical belt of dense water vapor which efficiently absorbs THz radiation \cite{tropicalbelt}. In northern South America, previous studies of astronomical site testing in the visible range have shown that high nighttime humidity conditions make this region suitable only for educational observatories \cite{pinzon}. However, considering that millimeter/sub-millimeter observations need not be done during the night, unusually dry, clear-daytime-sky, high altitude regions in the northern Andes could be suitable candidates for a world-class millimeter-wave observatory. \\

Astronomical observations in the millimeter and sub-millimeter wavelength range require that atmospheric effects affecting absorption at these wavelengths are kept to a minimum \cite{southpole,radford2016,chajpwv}. The main factor contributing to the atmospheric opacity is water vapor, which very efficiently absorbs light in the THz range \cite{pardocerni,pardocerni2,cont350} due to a continuum absorption spectrum formed by collisionally broadened absorption lines of water vapor in this frequency range  \cite{linecont,submmlines,turner2009}. In order to characterize a site according to its atmospheric transparency to THz radiation, it is necessary to retrieve the precipitable water vapor (pwv) profile using remote sensing techniques such as microwave radiometry \cite{radiometer,paine2000fourier,southpole2}, satellite measurements \cite{aqua,spacemols,suen2014,spaceradio}, radiosonde humidity measurements \cite{radiosonde,radiosonde2}, or indirectly via models of in situ climatological measurements \cite{climatology} or GPS-delay studies \cite{gpsmet,gpsorig,gpsmap}. \\

Most of the precipitable water vapor that can potentially affect the path of a THz photon hoping to get through the atmosphere actually exists in the troposphere. The lower the elevation of a potential site, the longer the path becomes for that photon, increasing its chances of being absorbed by a water vapor molecule \cite{liebe1989mpm,linecont,lababs}. For this reason, the quality of a site is determined by the atmospheric water vapor column above the potential site \cite{mkradio,southpole3,he2012,atacama}. This water vapor column is usually expressed as the amount of precipitable water vapor, given in mm.\\

The water vapor column above a potential site can change rapidly (during the course of the day) or it can change seasonally \cite{arm2013,caumont2016}. This is why medium-to-long term monitoring of the atmosphere above a site is required before building an expensive mm/sub-mm wave radio telescope. However, even in best-case scenarios, earth-based telescopes will always be limited by atmospheric absorption of light specially at THz frequencies, where frequency windows of observations are few and in some cases, narrow \cite{arrlim,limits}.\\ 

\begin{figure}
\begin{center}
\includegraphics[scale=0.4]{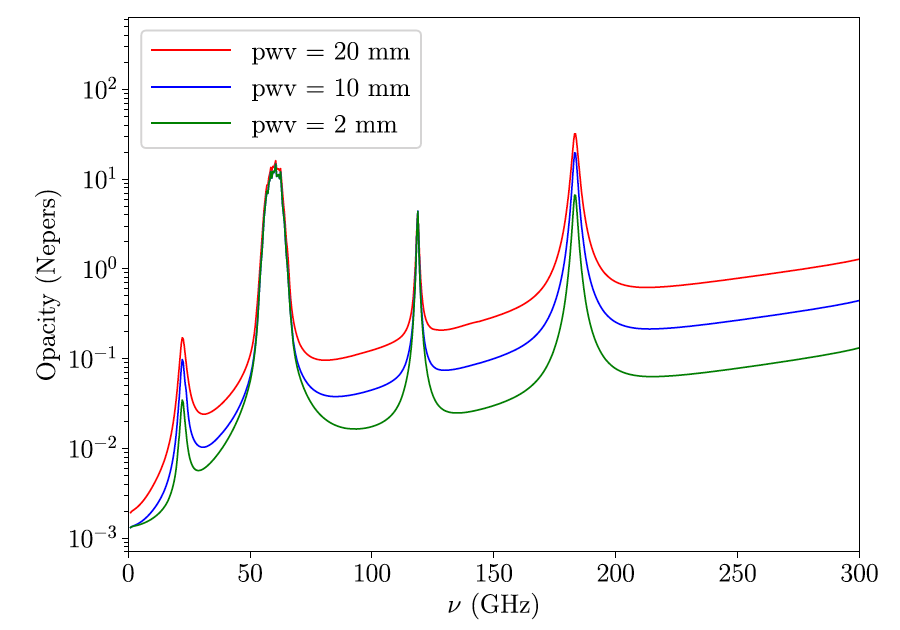}
\caption{Atmospheric opacity models for the Colombian Andes for different precipitable water vapor (pwv) values corresponding to three scenarios: \textit{Red line} (20 mm pwv) a 2600 masl site with 90\% relative humidity at 13$^\circ$C, \textit{Blue line} (10 mm pwv) a 3400 masl site with 90\% relative humidity at 0$^\circ$C, \textit{Green line} (2 mm pwv) a 3400 masl site with 30\% relative humidity at 0$^\circ$C.}\label{opac}
\end{center}
\end{figure}

We estimated the expected zenith opacity above the Colombian Andes (Figure \ref{opac}) at cm-mm wavelengths considering columns of 2, 10, and 20 mm of precipitable vater vapor and typical weather conditions. The elevation for the simulated sites in Figure \ref{opac} were chosen according to existing millimeter telescope sites (Pico Veleta, Plateau de Bure) \cite{mminf}. Thus, considering a best-case scenario (below 2 mm pwv), the potential observing frequency bands would be $70-115$ GHz, $120-170$ GHz, and $206-223$ GHz. The opacity models were calculated using the CASA \cite{casa} \texttt{asap.opacity.model} class, and deriving the pwv value from the 22 GHz K-band zenith opacity \cite{deuber,sanpedromartir}. \\

Although radiometer measurements are desirable for directly measuring the atmospheric water vapor column, their development and/or deployment can be complex and expensive \cite{radiopro,receiver}. For this reason we screened historic climate data for evidence of local long-term clear sky, low humidity conditions as proxies for a locally low precipitable water vapor column. By low-dimensionally embedding the data, we are able to reduce the dimensionality of multi-annual monthly data from $N_\mathrm{dim}=12$ to $N_\mathrm{dim}\le3$ while preserving $>95$\% of the variance in the data. Thus, we can correlate unusually dry regions regardless of their geographical location. Given that the climatological variables were sparsely measured by weather stations across Colombia, we evaluated the aptness of a given location using a Bayesian probability-derived quality index. Finally, we geographically clustered our candidate locations in order to identify regions of interest. We compared these regions of interest with regions with unusually low precipitable water vapor in low-resolution ($\sim20$ km) water vapor satellite maps \cite{suen2016} and find a possibly significative geographical correlation.  In an upcoming paper we will analize satellite data in order to retrieve high-resolution seasonal precipitable water vapor maps for the regions of interest reported here.\\

This paper is organized as follows. We describe our dataset and the distribution and types of weather stations in Section 2. In Section 3 we describe how we applied low-dimensional embedding algorithms to our data. This is followed by a discussion in Section 4, where we include an overview of climate patterns in Colombia, and describe our Bayesian probabilistic quality index to assess the suitability of a given weather station. We summarize our results in Section 5, where we identify candidate regions of interest for a mm-wave astronomical observatory site. Finally, we discuss our main conclusions and future perspectives in Section 6 .

\section{The Dataset}

\begin{figure}
\begin{center}
\includegraphics[scale=0.5]{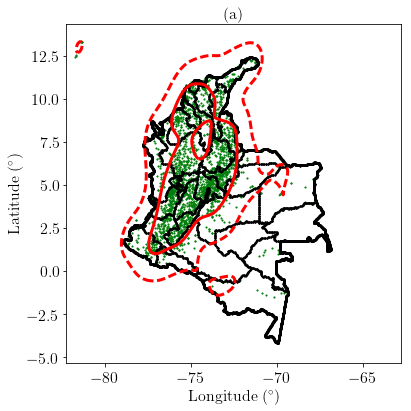}
\includegraphics[scale=0.5]{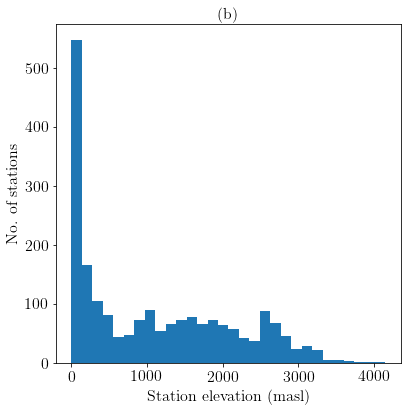}
\caption{Distribution of weather stations in the dataset. (a) 1$\sigma$ (solid) and 2$\sigma$ (dashed) countours of a gaussian kernel density estimation of the geographical distribution of weather stations (small green points) from the IDEAM 1981-2010 database. The estimation was done using the Python \texttt{scipy.stats.gaussian\_kde}  function \cite{scipy}. (b) Altitude histogram for all stations in the dataset.}\label{totmap}
\end{center}
\end{figure}

The meteorological data used in this study were gathered over 30 years (1981-2010) in 2046 weather stations monitored by the Instituto de Hidrolog\'ia, Meteorolog\'ia y Estudios Ambientales (IDEAM) in Colombia (Figure \ref{totmap}). The  variables that we considered relevant to our work were: Elevation (in meters above sea level, or masl), Precipitation (mm/mo), Rain Days (d/mo),  Relative Humidity (\%), and Sunshine Duration (h/d). The climatological variables are reported as multi-annual monthly data, i.e. each variable is reported monthly from January to December averaged over the 30 year range. For precipitation, each monthly datum corresponds to the cumulative monthly value.  For rain days, each monthly datum corresponds to the number of rain days for that month.  For relative humidity, each monthly datum corresponds to the  average daily value averaged over each month. For sunshine duration, each monthly datum corresponds to the daily value averaged over each month. All stations reported precipitation values, although only 2002 reported rain days, 445 reported relative humidity, and 336 reported sunshine duration. This means that not all stations registered all the meteorological variables relevant to this work. Thus, we classify stations according to which variables they measured (Table \ref{tabstype}). The data is public and can be found in the repository for this paper at \texttt{https://github.com/saint-germain/ideam} .\\

\begin{table}
\caption{\label{tabstype}Types of stations classified according to its reported variables. Y = measured, N = not measured.}
\begin{indented}
\item[]\begin{tabular}{@{}cccccc}
\br
Type&Precip.&Rain &Relative&Sunshine&No. of Stations\\
&&Days&Humidity&Duration&\\
\mr
$T_1$&Y&N&N&N&29\\
$T_2$&Y&Y&N&N&1563\\
$T_3$&Y&N&Y&N&1\\
$T_4$&Y&Y&Y&N&117\\
$T_5$&Y&N&N&Y&1\\
$T_6$&Y&Y&N&Y&8\\
$T_7$&Y&N&Y&Y&13\\
$T_8$&Y&Y&Y&Y&314\\
\br
\end{tabular}
\end{indented}
\end{table}

\section{Low-Dimensional Embedding of Climatological Data}

Our goal is to classify and group together stations that show a similar climatological behavior for each variable, regardless of the station's location, elevation or type. Since climate data track multi-annual seasonal variations, we can correlate weather patterns of regions which are not obviously related climatologically but might show a similar behavior, e.g. dry weather patterns in Guajira (a desert region in the north of Colombia) with similar dry conditions present at high-mountain sites in the Andes.\\

We classified stations using a low-dimensional embedding \cite{embed} of the available climate data. This ensures that a classification algorithm can uncover features in the data even when the data are projected across a number of dimensions that is lower than its native dimensionality (which in this case is monthly, i.e. $N_\mathrm{dim}=12$). This involves performing a dimensionality reduction algorithm followed by a clustering algorithm. Given that we wish to make a grouping of stations showing similar climatological patterns without making assumptions on the data, we decided to use unsupervised learning techniques.  In our case, we chose Principal Component Analysis (PCA) covering 95\% of the variance followed by a Gaussian Mixture Model selected by a low/locally minimum Bayesian Information Criterion (BIC). 

\subsection{Principal Component Analysis}

\begin{figure}
\begin{center}
\includegraphics[scale=0.8]{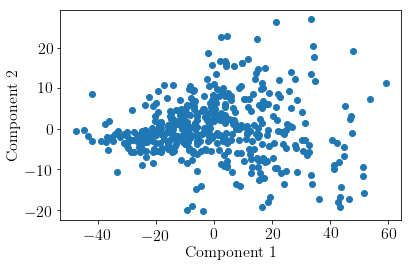}
\caption{Relative humidity data projected across 2 principal components. The dimensionality of the data has been reduced from 12 to 2 while covering 95\% of the variance of the data. }\label{pcah}
\end{center}
\end{figure}

We were able to reduce the dimensionality of the data for each variable from  $N_\mathrm{dim}=12$ to  $N_\mathrm{dim}^\mathrm{PCA}\le3$, while ensuring that at least 95\% of the variance of the data is explained by the least amount of components. For precipitation, rain days and sunshine duration, $N_\mathrm{dim}^\mathrm{PCA}(95\%)=3$ and for relative humidity $N_\mathrm{dim}^\mathrm{PCA}(95\%)=2$. To do this we used the Python \texttt{sklearn.decomposition.PCA} module \cite{sklearn}. To illustrate, Figure \ref{pcah} shows the relative humidity data projected across 2 principal components explaining 95\% of the variance, thus preserving most of the original structure of the data. Dimensionality-reduced precipitation, rain days, and sunshine duration data are shown later (Figure \ref{lde}).

\subsection{Gaussian Mixture Models}

After projecting the data along the components found by the PCA algorithm, we clustered the dimensionally-reduced data using a Gaussian Mixture Model, which uses an Expectation-Maximization algorithm to find a Maximum Likelihood model composed of a number $N_\mathrm{G-C}$ of Gaussian distributions, each of which represents a cluster in the data \cite{gmm}. Depending on the restrictions put on the covariance of each Gaussian distribution across the data space, the number of free parameters can go from $N_\mathrm{G-C}(N_\mathrm{dim}+1)$ (spherical covariance) to  $N_\mathrm{G-C}(3N_\mathrm{dim}+1)$ (full covariance). Since the number of components and covariance restriction is given by the user and not \emph{a posteriori} by the algorithm, we need to make sure that the Gaussian components do not over-fit the data, e.g. a model which yields one cluster per datum should be disallowed.\\

\begin{figure}
\begin{center}
\includegraphics[scale=0.5]{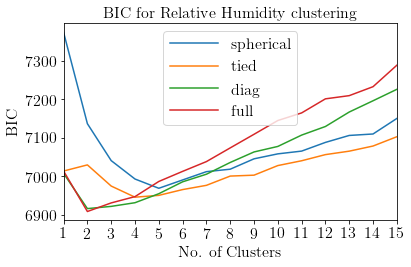}
\includegraphics[scale=0.5]{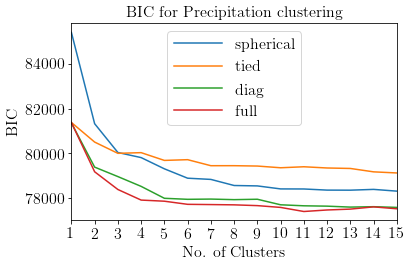}\\
\includegraphics[scale=0.5]{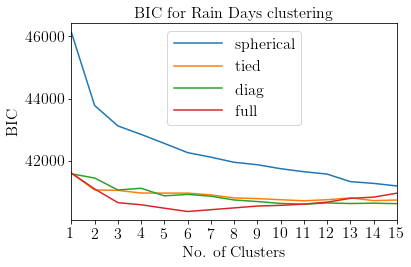}
\includegraphics[scale=0.5]{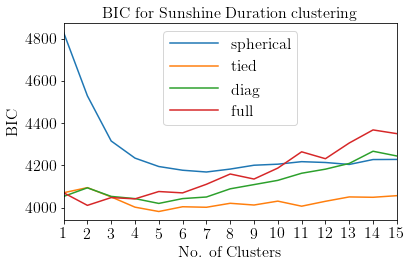}
\caption{Bayesian Information Criterion (BIC) results for Gaussian Mixture Models using a different number of clusters and using different covariance restriction methods \cite{sklearn} applied on each measured variable dataset.}\label{bic}
\end{center}
\end{figure}

\begin{figure}
\begin{center}
\includegraphics[scale=0.45]{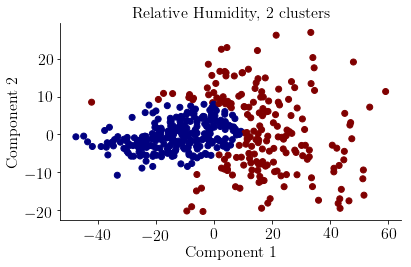}
\includegraphics[scale=0.45]{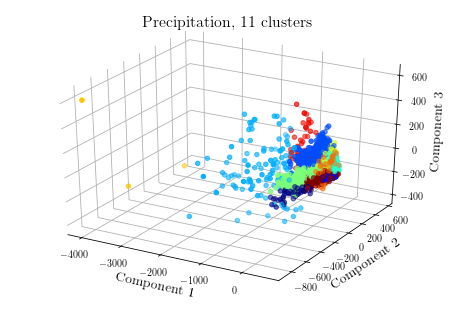}
\includegraphics[scale=0.45]{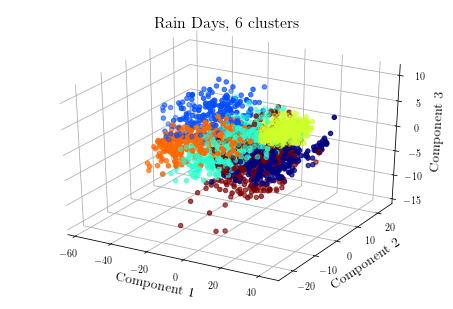}
\includegraphics[scale=0.45]{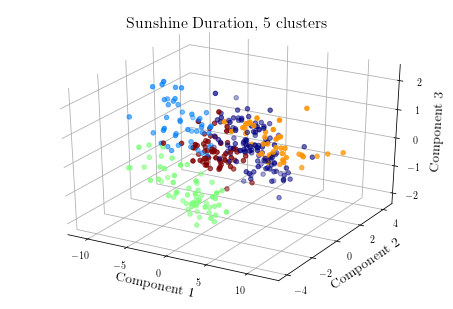}
\caption{Lowest BIC Gaussian Mixture Model clustering of reduced dimensionality climate data. Colors indicate the results of the clustering classification.}\label{lde}
\end{center}
\end{figure}

In order to achieve this, we compare the Bayesian Information Criterion \cite{bicref} for a grid of Gaussian clusters ($N_\mathrm{G-C}<20$) and covariance restriction methods for each variable (Figure \ref{bic}). To do this we used the Python \texttt{sklearn.mixture.GMM} module \cite{sklearn}. We selected the models which produced the lowest BIC value, and plotted the GMM-classified, PCA-projected climate data in Figure \ref{lde}. \\

\begin{figure}
\begin{center}
\includegraphics[scale=0.5]{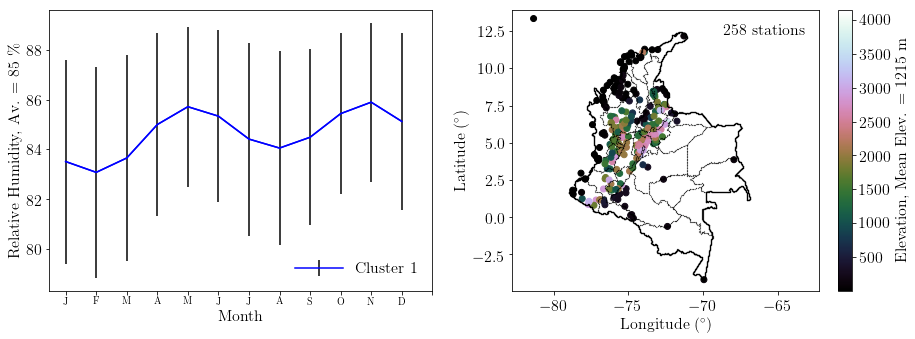}
\includegraphics[scale=0.5]{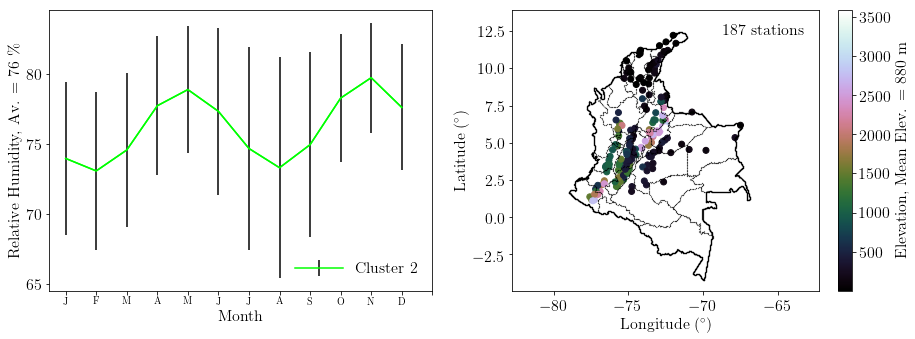}
\caption{Climate clusters (1-2 of 2) as predicted by the lowest-BIC Gaussian Mixture Model for relative humidity. Left column: Average monthly relative humidity and standard deviation (error bars) for each climate cluster. Right column: Geographical location and altitude of weather stations in each relative humidity cluster.}\label{clusth}
\end{center}
\end{figure}

\begin{figure}
\begin{center}
\includegraphics[scale=0.5]{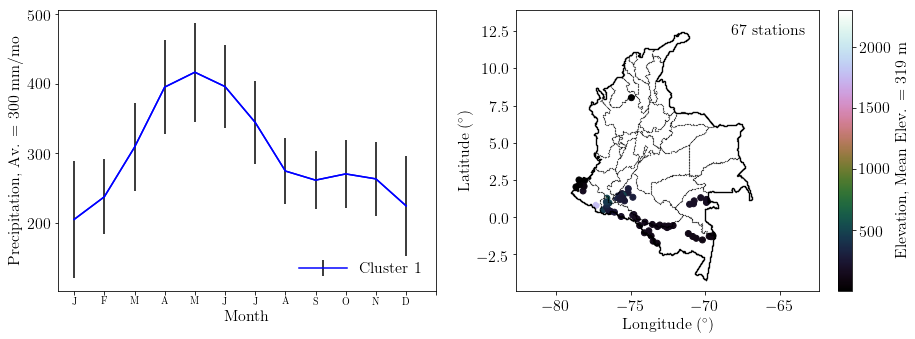}
\includegraphics[scale=0.5]{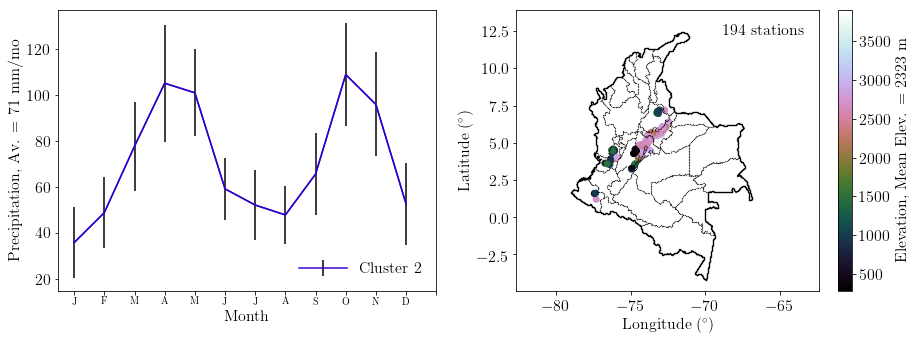}
\includegraphics[scale=0.5]{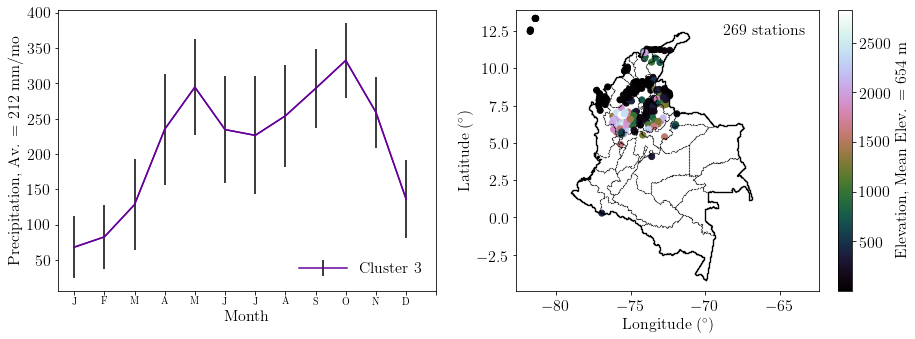}
\caption{Climate clusters (1-3 of 11) as predicted by the lowest-BIC Gaussian Mixture Model for precipitation. Left column: Average monthly precipitation and standard deviation (error bars) for each climate cluster. Right column: Geographical location and altitude of weather stations in each precipitation cluster.}\label{clustl}
\end{center}
\end{figure}

\begin{figure}
\begin{center}
\includegraphics[scale=0.5]{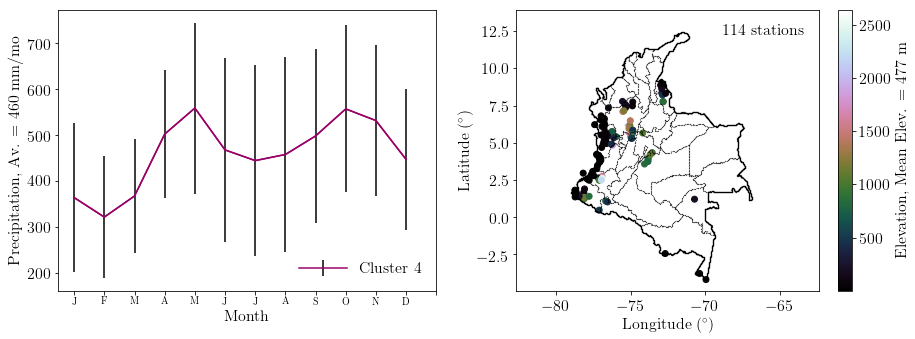}
\includegraphics[scale=0.5]{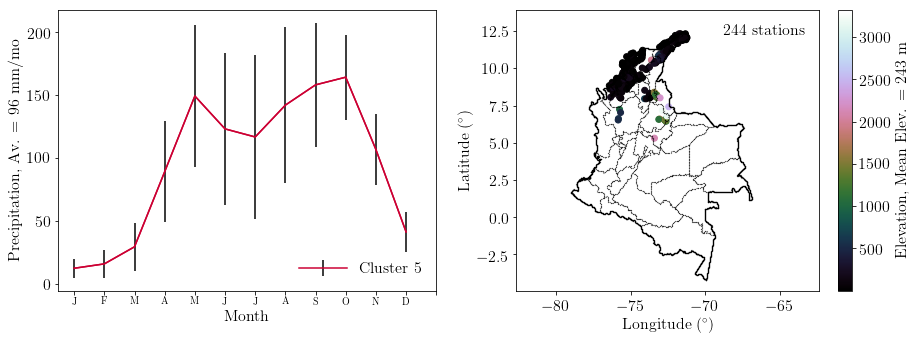}
\includegraphics[scale=0.5]{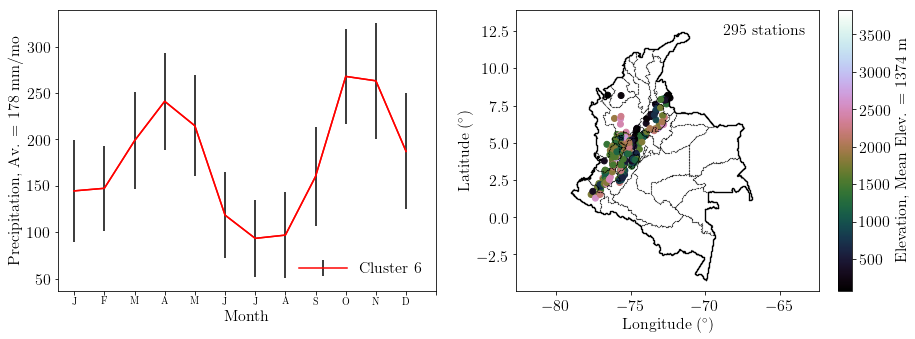}
\caption{Climate clusters  (4-6 of 11) as predicted by the lowest-BIC Gaussian Mixture Model for precipitation. Left column: Average monthly precipitation and standard deviation (error bars) for each climate cluster. Right column: Geographical location and altitude of weather stations in each precipitation cluster. Continuation of Figure \ref{clustl}.}\label{clustl2}
\end{center}
\end{figure}

\begin{figure}
\begin{center}

\includegraphics[scale=0.5]{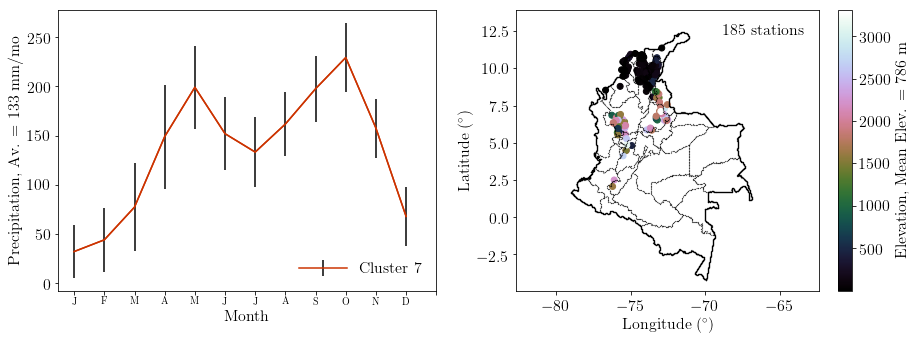}
\includegraphics[scale=0.5]{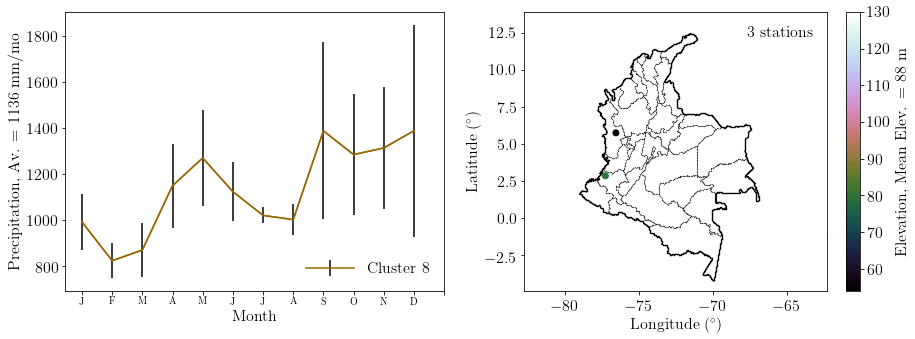}
\includegraphics[scale=0.5]{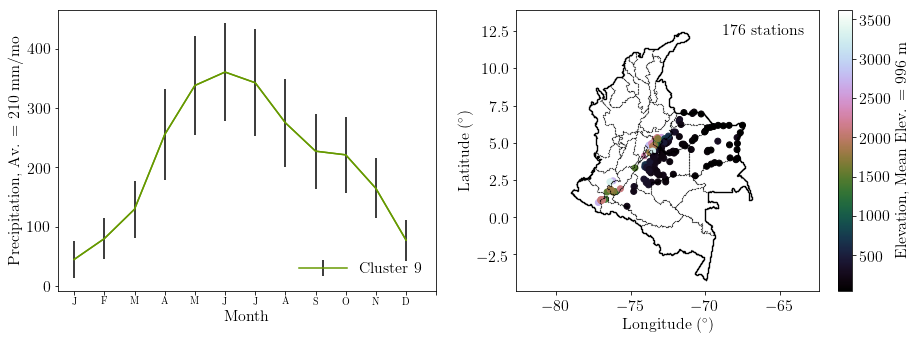}
\caption{Climate clusters (7-9 of 11) as predicted by the lowest-BIC Gaussian Mixture Model for precipitation. Left column: Average monthly precipitation and standard deviation (error bars) for each climate cluster. Right column: Geographical location and altitude of weather stations in each precipitation cluster. Continuation of Figure \ref{clustl2}.}\label{clustl3}
\end{center}
\end{figure}

\begin{figure}
\begin{center}
\includegraphics[scale=0.5]{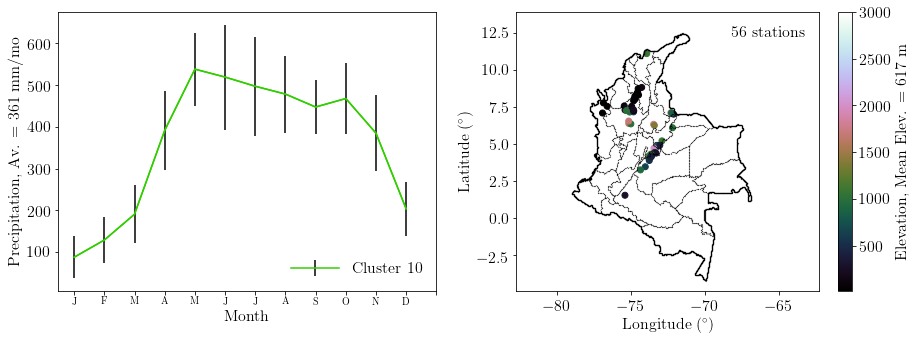}
\includegraphics[scale=0.5]{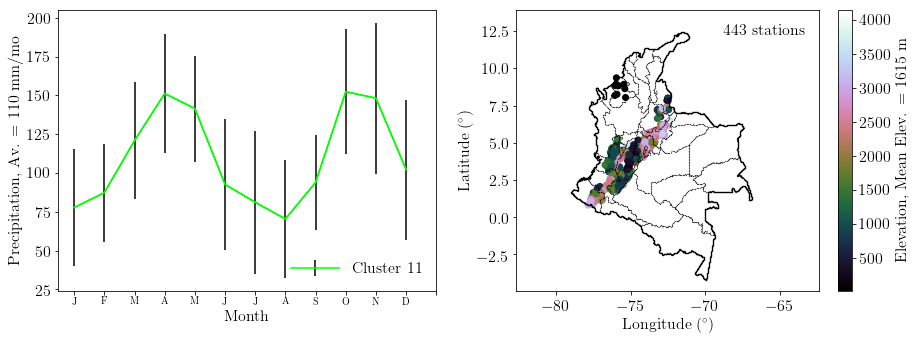}
\caption{Climate clusters (10-11 of 11) as predicted by the lowest-BIC Gaussian Mixture Model for precipitation. Left column: Average monthly precipitation and standard deviation (error bars) for each climate cluster. Right column: Geographical location and altitude of weather stations in each precipitation cluster. Continuation of Figure \ref{clustl3}.}\label{clustl4}
\end{center}
\end{figure}

\begin{figure}
\begin{center}
\includegraphics[scale=0.5]{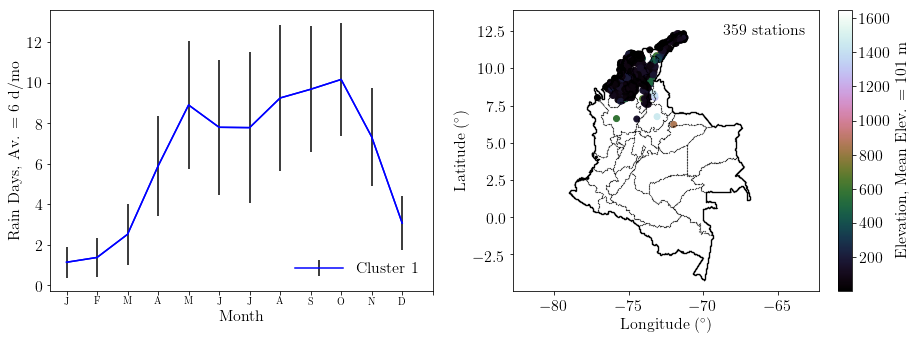}
\includegraphics[scale=0.5]{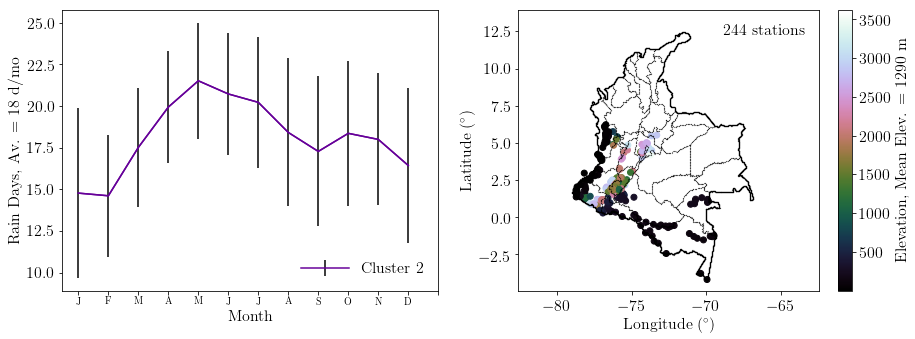}
\includegraphics[scale=0.5]{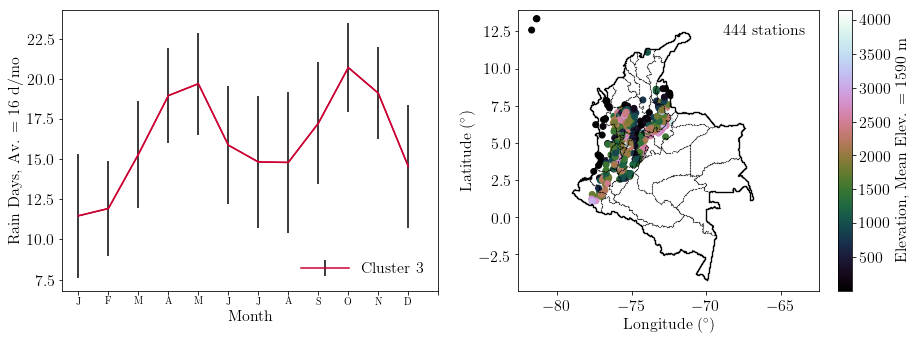}
\caption{Climate clusters (1-3 of 6) as predicted by the lowest-BIC Gaussian Mixture Model for rain days. Left column: Average monthly rain days and standard deviation (error bars) for each climate cluster. Right column: Geographical location and altitude of weather stations in each rain days cluster.}\label{clustd}
\end{center}
\end{figure}

\begin{figure}
\begin{center}
\includegraphics[scale=0.5]{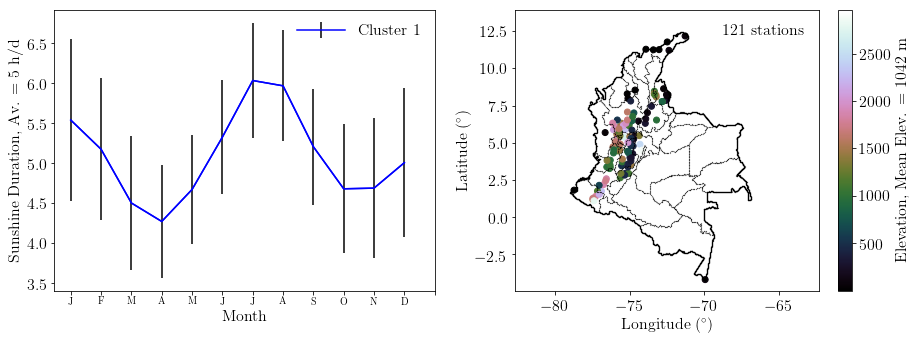}
\includegraphics[scale=0.5]{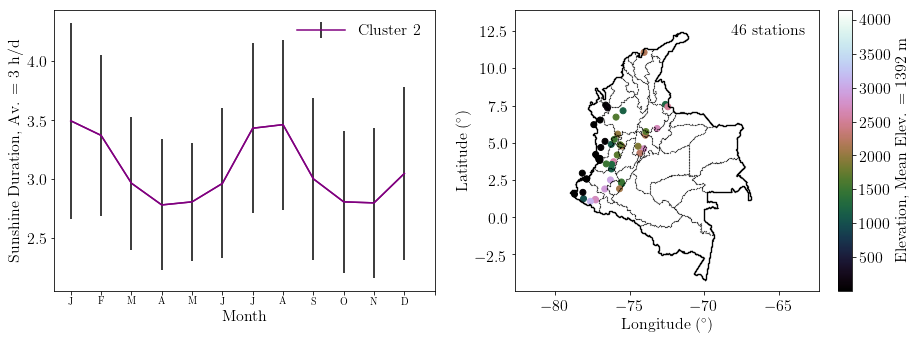}
\includegraphics[scale=0.5]{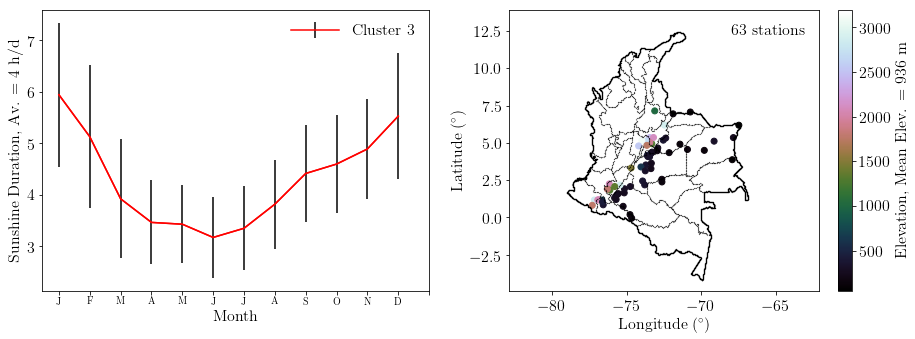}
\caption{Climate clusters (1-3 of 5) as predicted by the lowest-BIC Gaussian Mixture Model for sunshine duration. Left column: Average monthly sunshine duration and standard deviation (error bars) for each climate cluster. Right column: Geographical location and altitude of weather stations in each sunshine duration cluster.}\label{clustb1}
\end{center}
\end{figure}

\begin{figure}
\begin{center}
\includegraphics[scale=0.5]{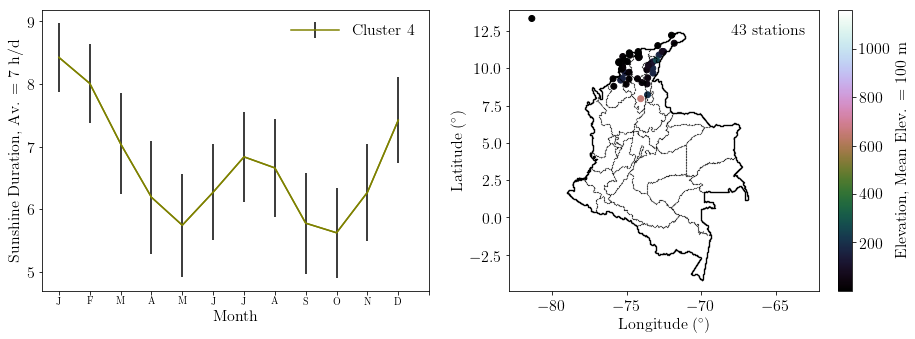}
\includegraphics[scale=0.5]{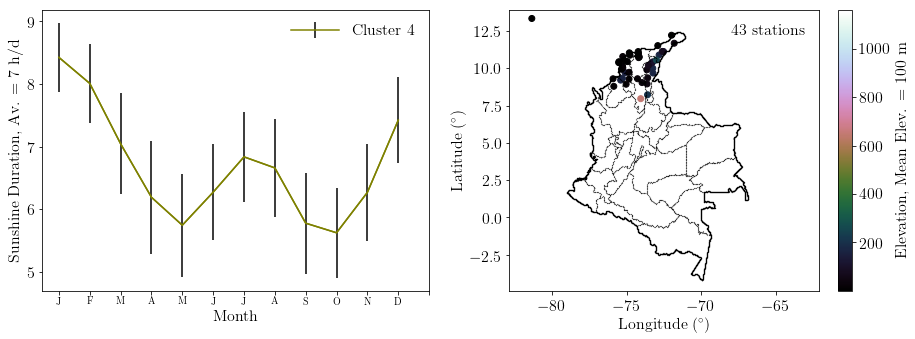}
\caption{Climate clusters (4-5 of 5) as predicted by the lowest-BIC Gaussian Mixture Model for sunshine duration. Left column: Average monthly sunshine duration and standard deviation (error bars) for each climate cluster. Right column: Geographical location and altitude of weather stations in each sunshine duration cluster. Continuation of Figure \ref{clustb1}.}\label{clustb}
\end{center}
\end{figure}

From the climatological clusters in Figures \ref{clusth}-\ref{clustb} we selected clusters that indicate a clear-sky, potentially dry climate. Table \ref{tabclu} shows the lowest-BIC number of clusters and covariance method for each variable along with our selected clusters. From here on, if a station appears in one of our preferred clusters, we will say that it satisfies our criterion for that specific variable.\\

\begin{figure}
\begin{center}
\includegraphics[scale=0.5]{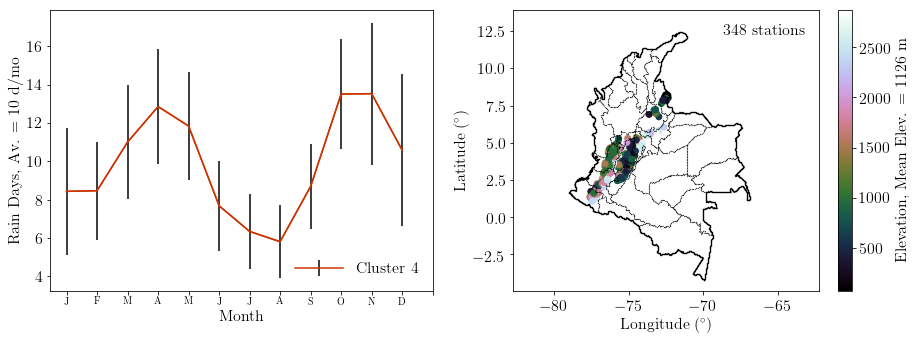}
\includegraphics[scale=0.5]{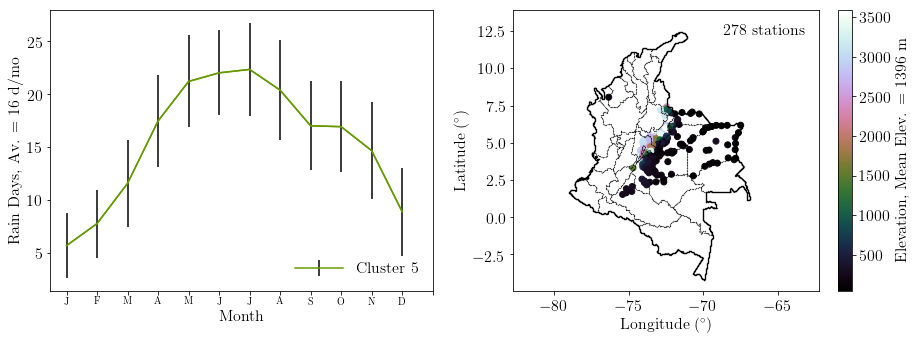}
\includegraphics[scale=0.5]{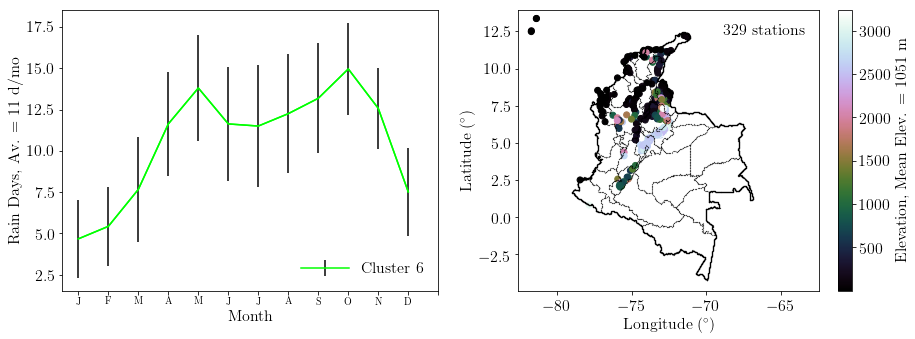}
\caption{Climate clusters (4-6 of 6) as predicted by the lowest-BIC Gaussian Mixture Model for rain days. Left column: Average monthly rain days and standard deviation (error bars) for each climate cluster. Right column: Geographical location and altitude of weather stations in each rain days cluster. Continuation of Figure \ref{clustd}.}
\end{center}
\end{figure}

\begin{table}
\caption{\label{tabclu}Climate clusters according to a lowest BIC-based selection of Gaussian Mixture Models following Figure \ref{bic}. The selected clusters column refers to the climate clusters shown in Figures \ref{clusth} to \ref{clustd}. }
\begin{indented}
\item[]\begin{tabular}{@{}lclll}
\br
Variable&Symbol& No. of&Covariance&Selected \\
&&clusters&Method &Clusters\\
\mr
Relative Humidity&$H$&2&Full&2\\
Precipitation &$R$&11&Full&2,3,7,11\\
Rain Days&$D$&6&Full&1,4,6\\
Sunshine Duration&$S$&5&Tied&1,4,5\\
\br
\end{tabular}
\end{indented}
\end{table}

\section{Discussion}

We wish to identify all stations for whom all four criteria (low humidity, precipitation, rain days, and high sunshine duration) for identifying a region as having a clear-sky, potentially dry climate can be met.  However, since only 6\% of the stations measured all variables (type $T_8$ in Table \ref{tabstype}), it is not straightforward to accept or reject stations based on their belonging to one of our selected clusters. In other words, a station that meets less than all four criteria can still be considered in our analysis, but it will not be given the same level of importance as a station that meets all four criteria. This requires setting up a method to determine the quality of a site depending on its location and station type.  Thus we selected stations that meet criteria for (and only for) the variables they measured. Thus, if a station measured precipitation and humidity (Type $T_3$), it had to belong to a precipitation and humidity cluster listed in Table \ref{tabclu} or else it was rejected. 665 stations across Colombia showed such behavior. \\

In order to estimate how much consideration we should give to a shortlisted station with an incomplete measurement set, we propose the use of a quality index based on a probabilistic analysis of the entire dataset, accounting for the non-uniform joint distribution of station types (Table \ref{tabstype}) and elevations (Figure \ref{totmap}b). Neglecting to include all stations in the calculation of this joint probability leads to very strong selection biases, assigning unfairly high quality indices to substandard stations. Therefore, even though we will later limit our selection to high-elevation sites in order to achieve lower pwv regions, we considered all the information provided by the full dataset in our probabilistic analysis. 

\subsection{Bayesian probabilistic analysis}\label{bpa}

We quantify the quality of a given site with the probability for the hypothesis that a station $j$ which appears in our shortlist meets all four criteria, given that it is located at a given elevation and is of a given type. We can write down this probability $P_j$ in terms of the following events:

\begin{itemize}
\item $C$ is the event that a given station meets all four criteria, i.e. belongs to our preferred clusters for the four variables relevant to this study.
\begin{itemize}
\item $C$ is actually the conjunction of all four of the $R,D,H,S$ events ``the station belongs to a selected cluster for variable $X$'', i.e. $C=R\cap D\cap H \cap S$ (See Table \ref{tabclu}).
\end{itemize}
\item $A$ is the event that a given station appears in our shortlist.
\item $T_i$ is the event that a given station is of type $T_i$ (See Table \ref{tabstype}).
\begin{itemize}
\item For example, if $M_X$ is the event ``the station measured variable $X$'', then the event $T_7$ is equivalent to $M_R\cap \neg M_D\cap M_H \cap M_S$
\end{itemize}
\item $h$ is the event that a given station is located at an elevation $h$.
\end{itemize}

Thus, the probability that a station $j=\{0,1,...,N_\mathrm{shortlist}\}$ of type $T_i$, meets all four criteria ($C$) and appears in our shortlist ($A$) given that it is located at an elevation $h$, can be written as $P_j=P(C\cap A\cap T_i\ |\ h)$. We cannot  calculate exactly this probability, as it depends on unknown unknowns, i.e. on whether a station that did not measure a given variable in reality belongs to one of our preferred climatological clusters for that variable (Table \ref{tabclu}). Thus we will use Bayes' theorem to help estimate this probability using simpler conditional probabilities. First we control for the inhomogeneous distribution of elevations $P(h)$,
\begin{equation}
P_j=P(C\cap A\cap T_i\ |\ h)=\frac{P(C\cap A\cap T_i\cap h)}{P(h)}\ .
\end{equation}
The joint probability of all events $C\cap A\cap T_i\cap h$ can be rewritten as,
\begin{equation}
P_j=\frac{P(h\ |\ C\cap A\cap T_i)}{P(h)}P(C\cap A\cap T_i)\ .
\end{equation}
If a station satisfies all four criteria, $P(C\cap A\cap T_i)=1$. This probability can be expressed in terms of simpler conditional probabilities,
\begin{equation}\label{peejay}
P_j=\frac{P(h\ |\ C\cap A\cap T_i)}{P(h)}P(C\ |\ A\cap T_i)P(A\ |\ T_i)P(T_i)\ ,
\end{equation}
where 
\begin{itemize}
\item $P(h\ |\ C\cap A\cap T_i)$ is the distribution of elevations for a station type ($T_i$) on our shortlist ($C\cap A$), 
\item $P(A\ |\ T_i)$ is the probability that a station of type $T_i$ is in our shortlist, 
\item $P(T_i)$ is the probability that a station type is $T_i$, and 
\item $P(C\ |\ A\cap T_i)$ is the probability that a station meets all four criteria ($C$) given that it is in our shortlist ($A$) and is of type $T_i$. 
\end{itemize}
All of these probabilities except for the last one can be computed directly from the data\footnote{Distributions of elevation-related probabilities were approximated discretely using a bin size estimated from the mean elevation difference between stations when sorted by elevation. }.\\

\begin{equation}\label{pclanti}
P(C\ |\ A\cap T_i)=\frac{P(C\cap A\cap T_i)}{P(A\cap T_i)}\  .
\end{equation}
In order to estimate $P(C\ |\ A\cap T_i)$ we assume that the probability of a station meeting our criteria for a given number of variables is independent of the station type. To illustrate this, let us assume that a given station is of type 1, so it only measured precipitation and rain days, i.e.,
\begin{equation}
T_1=M_R\cap M_D\cap\neg M_H\cap\neg M_S\ .
\end{equation}
If this station of type $T_1$ is in our shortlist ($A$), the $R$ and $D$ criteria are already met, so $A\cap R \cap D=A$. Thus,
\begin{equation}
P(C\cap A\cap T_1)=P( H\cap S\ |\ A\cap T_1)\ .
\end{equation}
Substituting this into Eq. (\ref{pclanti}),
\begin{equation}
P(C\ |\ A\cap T_1)=\frac{P( H\cap S\ |\ A\cap T_1)}{P(A\cap T_1)}=P(H\cap S\ |\ A\cap T_1) \ .
\end{equation}
The last term in the previous expression is unknown, but we can approximate it by assuming that it is independent of the station type. Therefore, 
\begin{equation}
P(C\ |\ A\cap T_1)\simeq P(H\cap S)\ .
\end{equation}
The probability on the right hand side of the previous equation can be directly computed from the data, as it only requires counting how many stations meet both $H$ and $S$ criteria. This procedure can be extended to all other station types.\\

The probabilities for stations in our shortlist ($P_j$ in Equation \ref{peejay}) can be used to evaluate the quality of a site, but since they vary by orders of magnitude, for the sake of clarity we decided to use instead a probability-derived logarithmic quality index. Thus, we define the quality index $Q_j$ for a station $j$ in our shortlist as,
\begin{equation}
Q_j=9\ \frac{\log (\nicefrac{P_\mathrm{j}}{P_\mathrm{min}})}{\log (\nicefrac{P_\mathrm{max}}{P_\mathrm{min}})}+1\ .
\end{equation}
Here $P_\mathrm{min},P_\mathrm{max}$ are the lowest and highest values for the probabilities for the stations in the first shorlist of 665 stations that satisfy criteria for all measured variables. Thus, if $P_j=P_\mathrm{max}$, $Q_j=10$, and if $P_j=P_\mathrm{min}$, $Q_j=1$.

\subsection{Final selection}

\begin{figure}
\begin{center}
\includegraphics[scale=0.55]{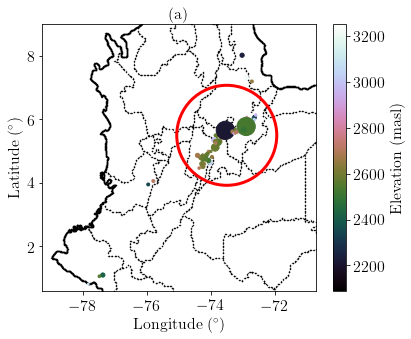}
\includegraphics[scale=0.55]{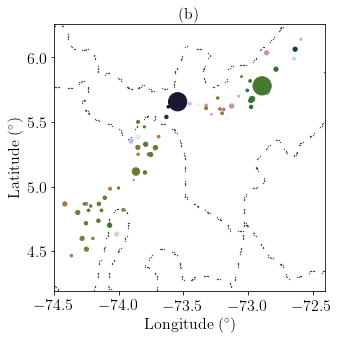}
\caption{Map of stations in our shortlist. The highlighted region in (a) is seen in more detail in (b), which corresponds approximately to the Colombian \emph{altiplano cundiboyacense} region. Marker size is proportional to the probability $P_j$, where the largest size corresponds to $Q_j=10$.}\label{shortlist}
\end{center}
\end{figure}

As mentioned above, as we go higher in elevation, the water vapor column decreases. For this reason we included an additional criterion for elevation limiting our list to 119 stations located above 2000 masl. This list was further reduced by requiring that for any given station, the total average plus 1$\sigma$ for precipitation, rain days, and relative humidity is respectively less than 4.3 mm/d, 17 d/mo, and 81\%, and for sunshine duration, more than 4 h/d. This left us with a shortlist of 83 weather stations.\\

Many of the stations in our dataset are lacking in relative humidity and sunshine duration data. In order to see if we could extrapolate using data from nearby stations, we estimated the difference between relative humidity and sunshine duration data for all stations in the entire dataset. For stations at a relative elevation of less than 100 m and located at less than 4 km away from each other, this difference amounts to less than 15\%. Thus, we extrapolated data for some of the 83 candidate stations in our shortlist, and were able to reject 4 stations which did not satisfy the relative humidity and sunshine duration criteria. The remaining 79 stations are plotted in Figure \ref{shortlist}a.\\

\begin{figure}
\begin{center}
\includegraphics[scale=0.55]{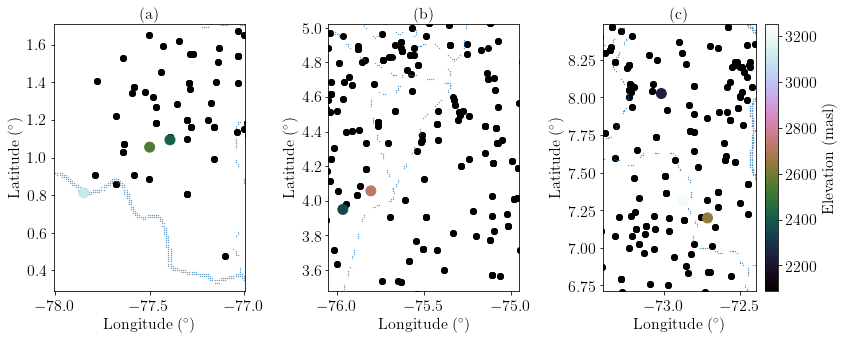}
\caption{Map of stations in our final shortlist but outside of the Colombian \emph{altiplano} region highlighted in Figure \ref{shortlist}. Marker size is proportional to the probability $P_j$, where the smallest size corresponds to $Q_j=2.2$.}\label{outliers}
\end{center}
\end{figure}

We rejected 9 additional stations located outside of the region located within the latitude, longitude range $[(4.3,6.2)^\circ,(-72.4,-74.5)^\circ]$ (Figure \ref{shortlist}a) due to geographical sparsity. Even though they could be indicative of clear-sky, potentially dry regions, there are not enough stations nearby to make any strong conclusions regarding those regions, as shown in Figure \ref{outliers}. Adding to this, in all 9 cases the only variable they measured was precipitation ($Q_j=2.2$), which led us to reject them from our shortlist. \\

The final 70 stations located in the Cundinamarca-Boyac\'a region of Colombia are shown in more detail in Figure \ref{shortlist}b. This figure shows two candidate locations in the northern region (Boyac\'a department), where the quality index $Q_j$ (proportional to the point size) indicates that all four criteria are satisfied. In order to see if those regions are geographically correlated, we grouped together these stations in a simplified manner using a lowest-BIC spherical covariance Gaussian Mixture Model on their coordinates (Figure \ref{unif}).

\section{Results}

We selected clusters of climatologically similar stations across Colombia using an unsupervised learning low-dimensional embedding algorithm for each measured variable. From these clusters, 70 weather stations came up as potential candidates for identifying high-mountain regions (at an elevation greater that 2000 masl) with clear-sky, potentially dry weather. The driest months of the year for all the candidate regions of interest are December-February and June-August (Figures \ref{clusth}-\ref{clustb}). It should be noted that  humidity in the Colombian Andes can change rapidly during the course of the day \cite{pinzon}, and daytime relative humidity measurements $<50$\% are often reported in high-mountain plateau stations.

\subsection{Candidate regions of interest}

\begin{figure}
\begin{center}
\includegraphics[scale=0.65]{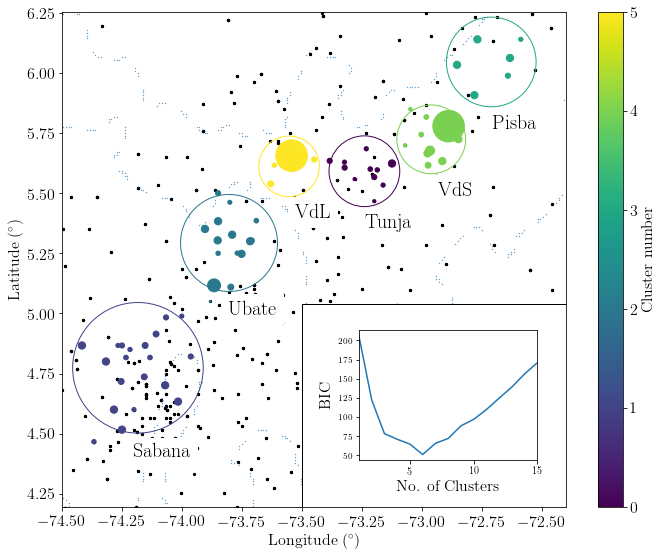}
\caption{Geographical clustering results  for a lowest-BIC spherical covariance Gaussian Mixture Model of the stations located inside the red circle in Figure \ref{shortlist}b, indicating the predicted variance regions for each cluster (red circles), and plotted along rejected stations in our original dataset (black points). Marker size is proportional to the probability $P_j$, where the largest size corresponds to  $Q_j=10$ and the smallest size corresponds to $Q_j=2.2$. Embedded plot: BIC results for different number of clusters. }\label{unif}
\end{center}
\end{figure}

Figure \ref{unif} shows 6 candidate regions of interest, where weather stations indicating clear-sky, potentially dry climate are regionally correlated within a $10-20$ km radius. In order to visualize the extent of those regions we plotted the predicted variance regions for each cluster along with other (rejected) stations in our sample in Figure \ref{unif}. It should be noted that the spherical variance GMM geographical clustering overestimates the extent of each region, and tells us nothing of their actual geographical shape. \\

We can now reject regions where the number of rejected stations within the predicted variance regions is similar or greater than the number of stations in our shortlist in Figure \ref{unif}.  We discuss the suitability of each of these regions below, from South-West to North-East based on our Bayesian probabilistic quality index (Section \ref{bpa}).

\subsubsection{La Sabana}

The first cluster is located at ($4.77^\circ$N, $74.18^\circ$W) and covers an radius of 15.3 km (Figure \ref{unif}, magenta circle). This is a region located to the west of Bogot\'a at an elevation above 2600 masl, with a large variety of microclimates. This explains the significant amount of stations from this region in our shortlist. However, Figure \ref{unif} shows that within this predicted variance region there are more stations that do not satisfy our criteria than stations that do, and no single sub-region can be identified. This, compounded with the fact that most (70\%) of these stations have a $Q$ value lower than 5.6 due to the lack of humidity/sunshine data, implies that at the moment this is not a strong candidate region of interest.

\subsubsection{Valle de Ubat\'e}

This cluster is located at ($5.30^\circ$N, $73.80^\circ$W) and covers a radius of 11.4 km (Figure \ref{unif}, cyan circle). This region, located to the NNE of Bogot\'a is a hilly region along a valley at an elevation above 2600 masl. The ratio of number of stations that do not meet our criteria to the stations in this predicted variance region is low (0.3), and almost half of these stations (46\%) have a $Q_j$ value higher than 6. This indicates a potential candidate region, although more  sunshine and humidity data are needed.

\subsubsection{Villa de Leyva}

This cluster is located at ($5.61^\circ$N, $73.55^\circ$W) and covers a radius of 7.11 km (Figure \ref{unif}, yellow circle). This region is located near the town of Villa de Leyva, and it is known for its dry weather.  The ratio of stations not in our shortlist vs. the stations in this predicted variance region is not very low (0.4). However, Figure \ref{unif} seems to indicate that the actual region of interest is narrower than the predicted variance region, as the stations that do not satisfy our criteria are on the NNW and SSW fringes of said region. The presence of one very high-$Q_j$ station and the narrow GMM-predicted variance is indicative of this region being a strong candidate. However, its relatively low elevation (near 2200 masl) can signify the presence of too much atmospheric water vapor above the surface.

\subsubsection{Cant\'on de Tunja}

This cluster is located at ($5.59^\circ$N, $73.24^\circ$W) and covers a radius of 8.29 km (Figure \ref{unif}, purple circle). The GMM-predicted variance region west of Tunja is located in the historic Tunja Canton, and even though the ratio of stations not in our shortlist to stations in this predicted region is low (0.31), there are simply not enough humidity and sunshine data to make any strong conclusion about this region. However, the high elevation of some of these stations is tantalizing, which is why we will keep this as a region of interest for our next paper.

\subsubsection{Valle del Sol}

This cluster is located at ($5.72^\circ$N, $72.96^\circ$W) and covers a radius of 8.06 km (Figure \ref{unif}, green circle). This is one of the most promising regions in our sample. Located in a wide, sunny valley (hence the name) at an elevation of 2600 masl, it is surrounded by mountains, and rains are not as common as elsewhere in the country. This region has the lowest ratio of not in our shortlist stations to stations in this predicted variance region (0.2), $40\%$ of the stations have a $Q_j$ value higher than 6, and one of the stations has a very high-$Q_j$ value. This will be one of the regions of interest for our next paper, and we think that it warrants radiometer measurements to be carried out.

\subsubsection{Parque Nacional Natural Pisba}

This cluster is located at ($6.04^\circ$N, $72.71^\circ$W) and covers a radius of 10.52 km (Figure \ref{unif}, teal circle). This is a \emph{p\'aramo} region, characterized by high-mountain tundra weather, with some mountains reaching an elevation of up to 3800 masl. The ratio of stations rejected stations to stations in this predicted variance region is high (0.5), and it is unlikely that the mist allows for a low-atmospheric water vapor region to be located here, even if rain is sporadic. The lack of a high-$Q$ station means that more sunshine-humidity data are needed, but this is possibly not a region of interest.

\subsection{Comparison to other works}

We compared our results to the work of  \citeasnoun{suen2014} and \citeasnoun{suen2016}, where they obtained global precipitable water vapor maps for 2012 using data from the MODIS instrument on board the Aqua and Terra satellites. Even though raw MODIS data has a potentially higher spatial resolution, the resolution of the processed maps is not fine enough to pinpoint a site location at a precision higher than $\sim20$ km. The area with the lowest mean precipitable water vapor that MODIS is able to measure ($\sim10-15$ mm) in the \citeasnoun{suen2014} map for Colombia spans the Tunja, Valle del Sol, Pisba, and part of the Valle de Ubat\'e regions. However, it also spans nearby regions with rejected stations, which means that low-resolution precipitable water vapor maps can indicate a wide, first order candidate region, but fail to account for highly local weather conditions, which are accounted for in our method. For this reason, in our next paper we will correlate the results presented here with existing GPS-delay water vapor measurements from the Colombian Global Navigation Satellite System GeoRED network (\texttt{https://geored.sgc.gov.co}) and seasonal, higher resolution MODIS data.

\section{Conclusions}

\begin{table}
\caption{\label{tabfin}Stations in two regions of interest showing a high-elevation daytime dry, clear-sky climate, organized by location, Bayesian probability quality measure ($Q_j$), distance to cluster center ($r_c$), mean precipitation  ($\bar{R}$), mean rain days ($\bar{D}$), mean relative humidity  ($\bar{H}$), mean sunshine duration  ($\bar{S}$), and elevation ($h$). Detailed table (including all clusters) available at \texttt{https://github.com/saint-germain/ideam} .}

\begin{tabular}{cllccccccc}
\toprule
Region of&   Station&                Municipality  &  $Q_j$ &  $r_c$  &   $\bar{R}$ &   $\bar{D}$  &   $\bar{H}$  &  $\bar{S}$ &  $h$ \\
interest &    code   &            &    &    (km) &   (mm/mo) &  (d/mo) &   (\%) &   (h/d) &  (masl) \\
\midrule
  &   21201410 &           Nemocon &  2.2 &            28 &       66 &       ND &    ND &  ND &    2600 \\
  &   24011060 &              Susa &  4.6 &            18 &       83 &       12 &    ND &  ND &    2600 \\
  &   24010140 &          Cucunuba &  4.7 &             6 &       57 &        9 &    ND &  ND &    2620 \\
  &   24015210 &         Sutatausa &  4.8 &             6 &       57 &       10 &    ND &  ND &    2700 \\
  &   24010170 &          Guacheta &  4.9 &            16 &       73 &       10 &    ND &  ND &    2690 \\
  &   24010070 &       Lenguazaque &  5.2 &            10 &       63 &       12 &    ND &  ND &    2650 \\
  &   24015130 &          Simijaca &  5.2 &            23 &       67 &       11 &    ND &  ND &    2572 \\
  &   21201620 &            Suesca &  5.6 &            20 &       59 &       10 &    ND &  ND &    2575 \\
                                       \rotatebox{90}{\rlap{Valle de Ubat\'e}}
  &   24010610 &  Carmen de Carupa &  6.1 &            12 &       54 &       12 &    ND &  ND &    2970 \\
  &   24011080 &          Cucunuba &  6.1 &             7 &       51 &        5 &    ND &  ND &    2562 \\
  &   24010280 &       Lenguazaque &  6.1 &             9 &       57 &       10 &    ND &  ND &    2585 \\
  &   24010440 &              Susa &  6.1 &            11 &       60 &        8 &    ND &  ND &    3130 \\
  &   24011090 &             Ubate &  6.1 &             4 &       62 &       13 &    ND &  ND &    2555 \\
  &   24015110 &             Ubate &  6.1 &             5 &       62 &       12 &    ND &  ND &    2610 \\
  &   21205400 &           Nemocon &  7.6 &            20 &       52 &       10 &    77 &   4 &    2580 \\
      \midrule
  &   24030760 &           Duitama &  4.0 &            17 &       69 &       10 &    ND &  ND &    2590 \\
  &   24030510 &             Paipa &  4.0 &            12 &       74 &        9 &    ND &  ND &    2900 \\
  &   24030790 &             Nobsa &  5.0 &             6 &       66 &       10 &    ND &  ND &    2500 \\
  &   24031040 &        Santa Rosa &  5.0 &            10 &       70 &        9 &    ND &  ND &    2500 \\
  &   24035140 &          Sogamoso &  5.0 &             4 &       59 &       10 &    ND &  ND &    2500 \\
  &   24030940 &          Sogamoso &  5.0 &             7 &       60 &       12 &    ND &  ND &    2500 \\
  &   24030410 &          Tibasosa &  5.0 &             5 &       62 &        7 &    ND &  ND &    2500 \\
                                      \rotatebox{90}{\rlap{~Valle del Sol}}
  &   24030410 &               Iza &  5.6 &            12 &       52 &        9 &    ND &  ND &    2470 \\
  &   24030540 &        Firavitoba &  6.1 &             6 &       57 &        9 &    ND &  ND &    2486 \\
  &   24030190 &            Mongui &  6.1 &            12 &       65 &        9 &    ND &  ND &    2970 \\
  &   24030760 &          Sogamoso &  6.1 &            11 &       65 &       10 &    ND &  ND &    3225 \\
  &   24035340 &          Sogamoso &  6.7 &             5 &       61 &       12 &    75 &   5 &    2500 \\
  &   24035150 &             Nobsa & 10.0 &            10 &       68 &       13 &    74 &   4 &    2530 \\

\bottomrule
\end{tabular}

\end{table}

In this paper we find plausible locations for a high-mountain cm- to mm-wave astronomical observatory in the northern Andes of Colombia, by analizing 30 years of climate data from 2046 weather stations. By low-dimensionally embedding the data, we are able to group together and correlate climate behavior indicative of daytime dry, clear-sky weather conditions.  A repository with all our code and the full dataset is available at \texttt{https://github.com/saint-germain/ideam}.\\

From our shortlist of 79 stations at elevations higher than 2000 masl that met our criteria, we selected 70 stations located  in the Cundinamarca-Boyac\'a \emph{altiplano} region of Colombia, which is a high-mountain basin with an average elevation of 2600 masl. Weather stations in our shortlist but outside this region are too sparsely located to suggest a candidate region of interest. \\

From those 70 stations we obtained 6 geographically correlated candidate regions of interest: La Sabana, Valle de Ubat\'e, Cant\'on de Tunja, Villa de Leyva, Valle del Sol, and Parque Nacional Natural Pisba. Seasonally, there appear to be two times of the year (Dec-Feb and Jun-Aug) where the weather conditions indicate the best daytime dry, clear-sky conditions. These months could be the best for mm-wave astronomical observations to be carried out. The Villa de Leyva region, despite having a dry, sunny climate, is located at a comparatively low elevation (2200 masl), which means that the precipitable water vapor column above this region is too high ($>20$ mm according to the \cite{suen2016} maps). The La Sabana region is also not a good candidate due to a high rejection rate of stations within this region.\\

We summarize the information about two regions of interest (Valle de Ubat\'e and Valle del Sol) in Table \ref{tabfin} with the highest Bayesian probabilistic indices, reflecting the quality of the data for stations in those regions. Most of the stations in this table are located at elevations above 2600 masl and below 3000 masl. We validated our results using satellite measurements of the upper bound for mean precipitable water vapor \cite{suen2014,suen2016} for those two regions, which is reported to be $\sim10-15$ mm in $\sim20$ km resolution maps. This means that it is not implausible to find a site with an even lower water vapor column in one of the regions of interest identified here. Some mountains nearby Valle de Ubat\'e and Valle del Sol have elevations of up to 3400 masl, and therefore could warrant further measurements using GPS stations, radio sondes and radiometers. The same could be said for the Pisba and Tunja regions, but the low quality of the climate data (indicated by low Bayesian probabilistic indices) indicates that more data is needed. For this reason, we will correlate the results for these four regions with GPS-delay water vapor measurements in our following paper.\\

There is an additional issue with the Valle del Sol region. The high industrial activity in this region fills the air with particulate material \cite{sogphd,sogamoso}, which would nullify the suitability of this region. Besides obtaining seasonal, higher-resolution precipitable water vapor maps, in our next paper we will study the amount of particulate material in the reported regions of interest using MODIS (Aqua and Terra) satellite data.\\
  
The methods used here can be extended and adapted to climatological datasets in other countries. Given a similar spatial and temporal coverage, our methods can provide a better picture of local climate than global satellite maps, and can help narrow down plausible locations for direct atmospheric opacity measurement campaigns. This can be a low-cost, preliminary step in radio astronomy site testing in developing countries.

\section*{Acknowledgments}

This work was made possible by the Vicerrector\'ia de Investigaci\'on 01-2015 research grant from Universidad ECCI. We would also like to thank the Instituto de Hidrolog\'ia, Meteorolog\'ia y Estudios Ambientales (IDEAM) for providing us with the dataset, and the anonymous referee for helping us convey our ideas more clearly.

 \section*{References}
 \bibliography{paperclimate}
\bibliographystyle{jphysicsB}
  
  \end{document}